\documentclass[
aps,%
12pt,%
final,%
notitlepage,%
oneside,%
onecolumn,%
nobibnotes,%
nofootinbib,
superscriptaddress,%
noshowpacs,%
centertags]%
{revtex4}
\usepackage[cp1251]{inputenc}
\usepackage[T2A]{fontenc}
\usepackage[english,russian]{babel}
\usepackage[english]{babel}
\usepackage{amsfonts}\usepackage{amsbsy}
\usepackage{feynmp}
\usepackage{graphicx}
\def\cal{\mathcal}
\newcommand{\di}{{\mathrm d}}

\newcommand{\Tr}{{\mathrm{Tr}}\,}
\newcommand{\ii}{{\mathrm i}}
\renewcommand{\Re}{{\mathrm{Re}}\,}
\renewcommand{\oint}{\int_{\cal C}}
\renewcommand{\Im}{{\mathrm{Im}}\,}
\def\Tc{{\cal T}_{\cal C}}
\def\scr#1{\mbox{\scriptsize #1}}

\def\vec#1{\mbox{\boldmath $#1$}}
\newcommand{\dpi}[1]{\frac{\di^4 #1}{(2\pi)^4}}
\newcommand{\Pbr}[1]{\left\{#1\right\}}
\newlength{\charwidth}
\def\medhat#1{\settowidth{\charwidth}{$#1\,$}{\makebox[\charwidth]{$\,
 {\widehat{\makebox[2mm]{$#1\,$}}}$}}\vphantom{#1}}
\newcommand{\lap}%
{\;\raisebox{-0.5ex}{$\stackrel{\scriptstyle <}{\scriptstyle \sim}$}\;}
\newcommand{\gap}%
{\raisebox{-0.5ex}{$\stackrel{\scriptstyle >}{\scriptstyle \sim}$}}

\def\lsim{\stackrel{\scriptstyle <}{\phantom{}_{\sim}}}
\def\gsim{\stackrel{\scriptstyle >}{\phantom{}_{\sim}}}

\def\Gr{G}\def\Se{\Sigma}

\def\A{A}
\def\Gm{\Gamma}
\def\F{F}
\def\Ft{\widetilde{F}}
\def\Ld{\Gamma^{\scr{out}}}
\def\Ldt{\Gamma^{\scr{in}}}
\def\Do{{\cal D}}

\def\contourxy{\unitlength=0.8cm
\begin{picture}(17,1)\thicklines
\contour
\put(14.5,-.5){\makebox(0,0){$\infty$}}
\put(11,-.5){\makebox(0,0){$t_x^+$}}
\put(8,1.8){\makebox(0,0){$t_y^-$}}
\put(11,0){\circle*{.2}}
\put(8,1){\circle*{.2}}
\end{picture}}
\def\contour{\thicklines
\put(1,-.5){\makebox(0,0){$t_{0}$}}
\put(16.5,.5){\makebox(0,0){$t$}}
\put(0,.5){\vector(1,0){16}}
\put(1,1.){\line(1,0){13}}\put(14,.5){\oval(1,1)[br]}
\put(14,0){\vector(-1,0){13}}\put(14,.5){\oval(1,1)[tr]}}
\linethickness{1.0pt}

\begin{document}


\selectlanguage{english}  
\title{
Nonlocal Form of Quantum Off-Shell Kinetic Equation\footnote{Dedicated to
S.T. Belyaev on the occasion of his 85th birthday.}}
\author{\firstname{Yu.B.~Ivanov}}
\email[]{Y.Ivanov@gsi.de}
\altaffiliation{RRC ``Kurchatov Institute'', Kurchatov sq.$\!$ 1, Moscow
123182, Russia}
\affiliation{Gesellschaft f\"ur Schwerionenforschung mbH, Planckstr.$\!$ 1,
64291 Darmstadt, Germany}
\author{\firstname{D.N. Voskresensky}}
\email[]{D.Voskresensky@gsi.de}
\altaffiliation{Moscow Institute for Physics and Engineering,
Kashirskoe sh. 31, Moscow 115409, Russia}
\affiliation{Gesellschaft f\"ur Schwerionenforschung mbH, Planckstr.$\!$ 1,
64291 Darmstadt, Germany}

\begin{abstract}
A new nonlocal form of the off-shell kinetic equation is derived.
While being equivalent to the Kadanoff--Baym and
Botermans--Malfliet formulations in the range of formal
applicability, it has certain advantages beyond this range. It
possesses more accurate conservation laws for Noether
quantities than those in the Botermans--Malfliet formulation. At
the same time the nonlocal form, similarly to the
Botermans--Malfliet one, allows application of the test-particle
method for its numerical solution, which makes it practical for
simulations of heavy-ion collisions.  The physical meaning of the
time-space nonlocality is clarified.
\end{abstract}
\maketitle
\today
\begin{fmffile}{cons-fig}

\section{Introduction}\label{Intr}

The work of S.T. Belyaev and G.I. Budker, who demonstrated the
Lorentz invariance of the relativistic distribution function and
derived relativistic Fokker--Planck kinetic equation \cite{BB56},
stands in the line of achievements of the kinetic theory. This
work entered  many textbooks and found numerous applications in
various fields. Presently the relativistic transport concepts are
a conventional tool to analyze the dynamics of dense and highly
excited matter produced in relativistic heavy-ion collisions.

A great progress was also 
achieved in microscopic foundation of the kinetic theory. The
appropriate frame for description of non-equilibrium processes
within the real-time formalism of quantum-field theory was
developed by Schwinger, Kadanoff, Baym and Keldysh
\cite{Schw,Kad62,Keld64}. The formalism allows extensions of the
quantum kinetic picture beyond conventional approximations (like
the quasiparticle one). This is caused by the quest for dynamical
treatment of broad resonances as well as stable  particles which
acquire a considerable mass width because of collisional
broadening. The above mentioned applications request for
development of approximate self-consistent schemes possessing
conservation laws being at least approximately 
satisfied \cite{Baym,IKV,IKV99,IKHV00,KIV01,IKV03,KRIV}.
Based on these schemes, numerical transport methods for treatment
of the off-shell dynamics have been developed
\cite{Mosel,Cass99,Leupold}.

Two slightly different forms of the Kadanoff--Baym equations
expanded up to first-order space-time gradients are now used: the
proper Kadanoff--Baym (KB) form, as it follows right after the
gradient expansion of exact KB equations \cite{Kad62}, and the
Botermans--Malfliet (BM) one, as it follows after a modification
of a Poisson-bracket term in the KB equation \cite{Bot90}. 
Both the KB and BM
forms coincide in the first-order gradient approximation but
differ in higher orders. Both forms have their advantages and
disadvantages \cite{IKV03}. The KB form possesses exact
conservation laws for the Noether current and the energy-momentum
\cite{KIV01,IKV03}, however, it does not allow the efficient
test-particle method to be applied to its numeric solution. The BM
form is very suitable for the test-particle method
\cite{Mosel,Cass99,Leupold} but only approximately conserves the
Noether current and the energy-momentum. 

In this paper we would like to put forward a new, space-time
nonlocal form of the quantum kinetic equation, which combines
favorable features of the above mentioned KB and BM forms. In
sect. \ref{sect-Kin-EqT} we start with brief review of properties
of the KB and BM forms of the off-shell kinetic schemes. Technical
details are deferred to the Appendix \ref{Contour}. In sect.
\ref{New} the new nonlocal form of the quantum kinetic equation is
derived and  physical meaning of the nonlocality  is clarified.
Possible applications  in numerical transport schemes are
discussed. 

\section{Kadanoff--Baym and Botermans--Malfliet Kinetics}
\label{sect-Kin-EqT}

In this section we summarize the formulation of the off-shell
kinetic equations in the two different forms: in the KB form
and in the BM form.
We assume the reader is familiar with the real-time
formulation of non-equilibrium many-body theory and use of the
contour matrix notation, detailed in Appendix \ref{Contour}.

Starting point of all considerations is the set of Kadanoff-Baym
equations which express the space-time changes of the Wigner
transformed\footnote{The space-time variable is
$X\equiv X^{\mu}\equiv   (t,{\bf x})$, and the
Fourier transformed variable is $p\equiv p^{\mu}$.
}
correlation function $\ii \Gr^{-+}(X,p)$ in terms of the
real-time contour convolution of the self-energy $\Se$ with the Green
function $\Gr$. We give the kinetic equation in compact notation (cf. Eq.
(\ref{H=FG}))
%
\begin{eqnarray}
\label{KB-eq.}
v_{\mu}\partial^{\mu}_X\ii\Gr^{-+}(X,p) =
\left[ \Se\otimes\Gr - \Gr\otimes\Se \right]^{-+}_{X,p}
\quad\mbox{with}\quad
v^\mu=\frac{\partial}{\partial p_{\mu}}G_0^{-1}(p),
\end{eqnarray}
%
where $\Gr^{-1}_{0}(p)$ is the Fourier transform of the inverse free Green
function
\begin{eqnarray}
\label{G0}
\Gr^{-1}_{0}(p)=\left\{
\begin{array}{ll}
p^2-m^2\quad&\mbox{for relativistic bosons}\\
p_0-{\vec p}^2/(2m)\quad&\mbox{for non-rel. fermions or bosons.}
\end{array}\right.
\end{eqnarray}
For a complete definition, Eq. (\ref{KB-eq.}) has to be supplemented with
further equations, e.g., for the retarded Green function together with
the retarded relations (\ref{Fretarded}). If a system
under consideration is only slightly spatially inhomogeneous and
slowly evolving in time, a good approximation is provided by an
expansion up to first order in space--time gradients.  Then the
main problem to arrive at a proper kinetic equation consists in
accurately disentangling a rather complicated r.h.s. of Eq.
(\ref{KB-eq.}). 

\subsection{$\Phi$-derivable approximations}

In actual calculations one often uses approximations or truncation
schemes to the exact non-equilibrium theory, where 
conservation laws (such as charge and energy--momentum
conservations) and thermodynamic consistency of the transport
theory are not evident. It was shown \cite{Baym,IKV,IKV99} that
there exists a class of self-consistent approximations, called
$\Phi$ derivable approximations, which are conserving at the
expectation value level, i.e. they provide true Noether currents and a
conserved energy--momentum tensor,  
and at the same time thermodynamically consistent. In these schemes the
self-energies are self-consistently generated from a functional
$\Phi[\Gr]$ through the following variational procedure
\cite{IKV99}
%
\begin{eqnarray}\label{var-Phi-component}
-\ii
\Se_{ik}(X,p)=\mp\frac{\delta\ii\Phi[\Gr]}{\delta\ii \Gr^{ki}(X,p)}
\times \left\{
\begin{array}{ll}
2&\mbox{ for real fields}\\
1&\mbox{ for complex fields}
\end{array}\right. ,
\quad i,k\in\{-+\}.
\end{eqnarray}
%
The functional $\Phi[\Gr]$ specifies a truncation scheme. It consists of
a set of properly chosen closed two-particle irreducible diagrams, where lines
denote the self-consistent propagators $\Gr$, while vertices are bare. The
functional variation with respect to $\Gr$ diagrammatically implies an opening
of a propagator line of $\Phi$.

\subsection{Physical notation}

It is useful to eliminate the imaginary factors inherent in the
standard Green function formulation and introduce quantities which
are real and possess clear physical meaning. Thus instead of Green
functions $\Gr^{ij}(X,p)$ and self-energies $\Se^{ij}(X,p)$ with
$i,j\in\{-+\}$  in the Wigner representation we  use the kinetic
notation of Refs. \cite{IKV99,IKV03}. We define the generalized
distribution functions $\F$ and $\Ft$ in the 8-dimensional phase
space,
%
\begin{eqnarray}
\label{F} F (X,p)
 =  (\mp )\ii \Gr^{-+} (X,p) ,\quad
\Ft (X,p)
= \ii \Gr^{+-} (X,p) .
\end{eqnarray}
%
Here and below the upper sign corresponds to fermions,
while the lower sign, to bosons,
%
\begin{eqnarray}
\label{A}
 A (X,p) \equiv -2\Im \Gr^R (X,p) = \Ft \pm F
\end{eqnarray}
is the spectral function, and $\Gr^R$ is the retarded propagator.
The spectral function satisfies the sum rule
%
\begin{eqnarray}
\label{A-sumf-nr}
\int_{0}^{\infty} \frac{\di p_0}{2\pi}
A(X,p) &=& 1 \quad \mbox{for nonrelativ. particles},
\\
\int_{-\infty}^{\infty} \frac{\di p_0}{2\pi} p_0 A(X,p) &=& 1
\quad \mbox{for relativ. bosons},\nonumber
\end{eqnarray}
%
which follows from the canonical equal-time (anti)commutation relations
for (fermionic) bosonic field operators.
The gain and loss rates of the collision term are defined
as
%
\begin{eqnarray}
\label{gain}
\Ldt (X,p) =   \mp \ii \Se^{-+} (X,p),\quad
\Ld (X,p)  =  \ii \Se^{+-} (X,p)
\end{eqnarray}
%
with the damping width
%
\begin{eqnarray}
\label{G-def}
\Gm (X,p)&\equiv& -2\Im \Se^R (X,p) = \Ld (X,p)\pm\Ldt (X,p),
\end{eqnarray}
%
where $\Se^R$ is the retarded self-energy.

In terms of above kinetic notation, the gradient-expanded
Kadanoff--Baym equations are reduced to equations for real
quantities: for the real and imaginary parts of the retarded Green
function $\Gr^R$, and for the phase-space occupation $F$. Note
that the number of equations for $G^{ij}$ Green functions is four,
which twice exceeds the number of unknown functions ($F$ and
$G^R$).
Before the gradient expansion all equations were completely
consistent. However, after the gradient expansion their
interrelation is no longer obvious. Necessary interrelations have
been derived in Refs. \cite{IKV99,IKV03}.

The equations for the retarded propagator in the first-order
gradient approximation can be immediately solved with the result
\cite{Kad62,Bot90}
%
\begin{eqnarray}
\label{Asol} \Gr^R=\frac{1}{M(X,p)+\ii\Gm(X,p)/2}\Rightarrow
\left\{\begin{array}{rcl} A (X,p) &=&\displaystyle \frac{\Gm
(X,p)}{M^2 (X,p) + \Gm^2 (X,p) /4}\\[4mm] \Re\Gr^R (X,p) &=&
\displaystyle \frac{M (X,p)}{M^2 (X,p) + \Gm^2 (X,p) /4}
\end{array}\right.
\end{eqnarray}
%
with the  ``mass'' function
%
\begin{eqnarray}\label{M}
M(X,p)=\Gr^{-1}_{0}(p) -\Re\Se^R (X,p).
\end{eqnarray}
Note that the  algebraic solution (\ref{Asol})  is valid within
the  first-order gradient approximation, i.e. only the terms
$O(\partial_X^2)$ are omitted.

\subsection{Kadanoff--Baym form of kinetic equation}\label{KB}

In terms of above notation, the KB kinetic equation for $F$ in the
first-order gradient approximation takes the form, see
\cite{IKV99} for details,
%
\begin{eqnarray}
\label{keqk1}
\Do
F (X,p) -
\Pbr{\Ldt , \Re\Gr^R}
&=& C (X,p) .
\end{eqnarray}
%
We refer this as the {\em  kinetic equation in the
KB-choice}\footnote{If the system consists of  different particle
species, there is a set of coupled kinetic equations corresponding
to each species.}. Here $\Pbr{...,...}$ denotes the
four-dimensional Poisson bracket (\ref{[]}). The differential
drift operator is defined as
%
\begin{eqnarray}\label{Drift-O}
\Do = \left(v_{\mu}  - \frac{\partial \Re\Se^R}{\partial
p^{\mu}}\right)
\partial^{\mu}_X +
\frac{\partial \Re\Se^R}{\partial X^{\mu}} \frac{\partial
}{\partial p_{\mu}},
\end{eqnarray}
%
note that $\Do \psi = \Pbr{M , \psi}$ for
 arbitrary function $\psi$.
The collision term is defined as
%
\begin{eqnarray}
\label{Coll(kin)}
C
(X,p)&=&\Ldt (X,p) \Ft (X,p) - \Ld (X,p) \F
(X,p).
\end{eqnarray}
%
In terms of a  functional
$\Phi$ the explicit form of the collision term is
%
\begin{eqnarray}
\label{Coll-var}
C
(X,p) =
 \frac{\delta\ii\Phi
}{\delta\Ft(X,p)}\Ft(X,p)
-\frac{\delta\ii\Phi
}{\delta\F(X,p)}\F(X,p) ,
\end{eqnarray}
%
cf. Eq. (\ref{var-Phi-component}).

Within  the consistent gradient expansion
$ C\propto\partial_X$.
If the diagrams for the self-energy
contain internal vertices, it gives rise to  non-local effects in
the collision term
%
\begin{eqnarray}
\label{nlColl(kin)} C  = C_{\rm loc}+C_{\rm mem}.
\end{eqnarray}
%
Here $C_{\rm loc}$ is a local part of $C$, where all these
nonlocalities are disregarded, and $C_{\rm mem}\propto\partial_X$
is the nonlocal part expanded up to first gradient terms.
Following \cite{IKV99} we call these effects the memory effects.
Explicit form of the $C_{\rm mem}$ depends on the specific system
under consideration.

For the sake of clarity,  below we do few simplifications. We
do not consider the memory effects, implying that
$C=C^{\scr{loc}}$. Besides,
 we confine ourselves to the case void of derivative coupling. Also we
do not explicitly
  introduce the particle-specie label to avoid overcomplication of
  equations.

The local part of the collision term  is charge
and energy--momentum conserving by itself
%
\begin{eqnarray}
\label{C-conser} \mbox{Tr} \int \dpi{p} \left(
\begin{array}{lll}
e\\p^\mu
\end{array}
\right) C^{\scr{loc}} =0.
\end{eqnarray}
%
Here   $e$ denotes a charge (e.g., the baryon number), while
  $\Tr$ implies the sum over all possible internal degrees of freedom
(like spin, isospin, etc.) and over possible
particle species.

 The true Noether current of the charge $e$,
%
\begin{eqnarray}
\label{c-new-currentk} j^{\mu} (X) = e \mbox{Tr} \int \dpi{p}
v^{\mu} F (X,p),
\end{eqnarray}
%
is {\em{exactly conserved}}
\begin{eqnarray}
\label{KBcons}
\partial_{\mu}j^{\mu} (X) = 0,
\end{eqnarray}
which follows right from the operator expression for this quantity,
cf. Ref.  \cite{IKV99}.

On the other hand, the integration of the KB equation
(\ref{keqk1}) permits us to derive another, ``effective KB
current"
%
\begin{eqnarray}
\label{KB-eff-currentk} j^{\mu}_{\scr{KB-eff}} (X)  = e \mbox{Tr}
\int \dpi{p} \left[\left(v_{\mu}  - \frac{\partial
\Re\Se^R}{\partial p^{\mu}}\right)F
-
\Re\Gr^R \frac{\partial\Ldt}{\partial p_{\mu}}
\right]
\end{eqnarray}
%
which is {\em{exactly conserved.}}
{\em{ The equivalence of (\ref{c-new-currentk}) and
(\ref{KB-eff-currentk}) }} and thus the exact conservation of the
Noether current (\ref{c-new-currentk}) follows from the
corresponding invariance of the $\Phi$ functional, cf. Eq. (6.9)
in \cite{IKV}, which results in the  consistency relation
%
\begin{equation}
\label{invarJk} 
e \mbox{Tr} \int \dpi{p}
\left\{
-\partial_\mu \left[ \frac{\partial
\Re\Se^R}{\partial p^{\mu}}F
+
\Re\Gr^R \frac{\partial\Ldt}{\partial p_{\mu}}
\right]
+ C \right\} = 0 .
\end{equation}
%
%
%
Similar situation takes place for the
energy-momentum conservation.
Here we also rely on $\Phi$-derivable approximations which
provide the corresponding exact conservation.

The conserving feature is especially important for devising numerical
simulation codes based on this kinetic equation.  Indeed, if a
test-particle method is used, one should be sure that the number of
test particles is exactly conserved rather than approximately.
In the test-particle method the distribution function is
represented by an ensemble of test
particles as follows
%
\begin{eqnarray}
\label{test-part}
F (X,p) \sim \sum_i \delta^{(3)} \left({\bf x} - {\bf x}_i (t)\right)
\delta^{(4)} \left(p-p_i (t)\right),
\end{eqnarray}
%
where the $i$-sum runs over test particles. Then the $\Do F$ term
in Eq. (\ref{keqk1}) just corresponds to the classical motion of
these test particles subjected to forces inferred from $\Re
\Se^{R}$, while the collision term $C$ gives stochastic change of
test-particle's momenta, when their trajectories ``cross''. For a
direct application of this method, however, there is a particular
problem with the KB kinetic equation. The additional term, i.e.
the Poisson-bracket term $\Pbr{\Ldt,\Re\Gr^R}$, spoils this
simplistic picture, since derivatives acting on the distribution
function $F$ appear here only indirectly and thus cannot be
included in the collisionless propagation of test particles. This
problem, of course, does not prevent a direct solution of the KB
kinetic equation. For instance, one can apply well developed
lattice methods, which are, however, much more complicated and
time-consuming as compared to the test particle approach.

\subsection{Botermans--Malfliet form of kinetic equation}\label{BM}

As can be seen from Eqs. (\ref{F}), (\ref{G-def}) and
(\ref{Coll(kin)}), the gain rate $\Ldt$ differs from $F\Gamma/A$
only by corrections of the first order in the gradients
%
\begin{eqnarray}
\label{BM-subst} \Ldt = \Gm F/\A + C/\A = \Gm F/\A +O(\partial_X),
\end{eqnarray}
%
since $C \sim O(\partial_X)$.
This fact permits us to neglect the correction $O(\partial_X)$, as
in the kinetic equation it leads to terms of already second-order
in the gradients. Upon substitution $\Ldt=\Gm F/\A$,  proposed by
Botermans and Malfliet \cite{Bot90}, one arrives at the following
form of the kinetic equation, see \cite{IKV99} for details,
%
\begin{eqnarray}
\label{keqk}
\Do
F (X,p) -
\Pbr{\Gm\frac{F}{\A},\Re\Gr^R} &=& C (X,p),
\end{eqnarray}
%
which is still equivalent to the KB form within the first-order
gradient approximation (all terms $\propto O(\partial_X^2)$ are
now omitted). We call the so obtained Eq. (\ref{keqk}) {\em the
kinetic equation in BM-choice}. All KB-choice properties of Eq.
(\ref{keqk}) within a $\Phi$-derivable approximation also
transcribe to BM-choice through the substitution $\Ldt=\Gm F/\A$
in the consistency relation Eq. (\ref{invarJk}).

The BM equation {\em exactly} conserves the following
``effective BM current"
%
\begin{eqnarray}
\label{BM-eff-currentk}
j^{\mu}_{\scr{BM-eff}} (X) &=& e \mbox{Tr}
\int \dpi{p} \left[ v^{\mu} F (X,p) + \Re\Se^R  \frac{\partial
F}{\partial p_{\mu}}
-
\Re\Gr^R \frac{\partial(\Gm F/\A)}{\partial p_{\mu}}
\right]
\\ \nonumber
&=&e \mbox{Tr} \int \dpi{p}
\frac{\Gamma}{2}B^{\mu}F ,
\end{eqnarray}
%
where
%
\begin{eqnarray}\label{B-j}
B^{\mu} \equiv (B_0 , \vec{B}) = A \left[ \left(v^{\mu}  -
\frac{\partial \Re\Se^R}{\partial p_{\mu}}\right) - \frac{M}{
\Gamma}\frac{\partial \Gamma}{\partial p_\mu} \right]
\end{eqnarray}
%
is the flow spectral function introduced in \cite{IKV99}. The
0-component of the flow spectral function $B^{\mu}$  obeys the
same sum-rule as the spectral function $A$, cf. \cite{Weinhold}.
The effective current (\ref{BM-eff-currentk}) differs from the
true Noether current $j^{\mu}$ of Eq. (\ref{c-new-currentk}) and the
effective KB current (\ref{KB-eff-currentk}) in terms of the
order of $O(\partial_X)$, provided a $\Phi$-derivable
approximation is used for self-energies. Thus with the BM choice
the conservation laws of the Noether current
(\ref{c-new-currentk}) and the energy--momentum tensor are
 only approximately fulfilled.

The effective BM-current (\ref{BM-eff-currentk}) was used by S.
Leupold \cite{Leupold} as a basis for the construction of a
test-particle ansatz for numerical solution of the nonrelativistic BM
kinetic equation.
To automatically fulfill the effective current
conservation, the test-particle ansatz is introduced for the
combination
%
\begin{eqnarray}
\label{test-part-L} \frac{1}{2} \Gm B_0 F (X,p) \sim \sum_i
\delta^{(3)} \left({\bf x} - {\bf x}_i (t)\right) \delta^{(4)}
\left(p-p_i (t)\right),
\end{eqnarray}
%
rather than for the distribution function itself. Note that the
energy $p^0_i(t)$ of the test particle is an independent
coordinate, not restricted by a mass-shell condition. W. Cassing
and S. Juchem \cite{Cass99} used this test-particle ansatz in the
relativistic case.

The BM kinetic equation  (\ref{keqk}) together with ansatz
(\ref{test-part-L}) for the distribution function result in the
following set of equations for evolution of parameters of the test
particles between collisions
\begin{eqnarray}
\label{x-dot} {\bf {\dot x}}_i &=& \frac{1}{v_0
-\partial_{E_i}\Re\Se^R -(M/\Gamma)\partial_{E_i}\Gamma }
\left({\bf v}_i + \nabla_{p_i} \Re\Se^R
+(M/\Gamma)\nabla_{p_i}\Gamma\right),
\\%
\label{p-dot} {\bf {\dot p}}_i &=& \frac{1}{v_0
-\partial_{E_i}\Re\Se^R -(M/\Gamma)\partial_{E_i}\Gamma}
\left(\nabla_{x} \Re\Se^R +(M/\Gamma)\nabla_{x_i}\Gamma\right),
\\%
\label{e-dot} {\dot E}_i &=& \frac{1}{v_0 -\partial_{E_i}\Re\Se^R
-(M/\Gamma)\partial_{E_i}\Gamma} \left(\partial_{t} \Re\Se^R
+(M/\Gamma)\partial_{t}\Gamma\right).
\end{eqnarray}

These equations of motion, in particular, give the time evolution
of the mass term $M$, of a test particle \cite{Cass99,Leupold}
%
\begin{eqnarray}
\label{off-shellness}
\frac{\di M_i}{\di t}=
\frac{M_i}{\Gm_i}\frac{\di \Gm_i}{\di t},
\end{eqnarray}
%
the origin of which can be traced back to the additional term
$\Pbr{\Gm F/\A,\Re\Gr^R}$ in the BM equation (\ref{keqk}).
Here $M_i(t)= M[t,{\bf x}_i(t);E_i(t),{\bf p}_i(t)]$ measures an
``off-shellness'' of the test particle,
 and $\Gm_i(t)=\Gm[t,{\bf
x}_i(t);E_i(t),{\bf p}_i(t)]$. Equation of motion
(\ref{off-shellness}) yields $M_i =\alpha_i \Gamma_i$, where
$\alpha_i $ do not depend on time, and implies that once the width
drops in time the particles are driven towards the on-shell mass,
i.e. to $M=0$. This clarifies the meaning of the additional term
$\Pbr{\Gm F/\A,\Re\Gr^R}$ in the off-shell BM transport: it
provides the time evolution of the off-shellness.

The problem of not exactly conserving Noether charges within the BM
kinetics can be mitigated in certain cases. 
For instance, if we start from initial conditions defined for
particles in vacuum (where $\Sigma = 0$ and hence both charges
coincide) and end the evolution also in a very dilute state, the
Noether charge turns out to be conserved in the end of the
evolution. However, this is not the case with heavy-ion
collisions, where we start with two cold nuclei.

\section{Nonlocal form of Off-Shell Kinetic Equation} \label{New}

\subsection{Nonlocal kinetic equation and conservation laws}
Let us rewrite the KB kinetic equation (\ref{keqk1}) in  the
following non-local (NL) form
%
\begin{eqnarray}
\label{new-kin-eq}
&&\Do \F (X,p) - \Pbr{\Gm\frac{F}{\A},\Re\Gr^R}
=C^{\rm NL}
\cr
&&\equiv
\left(1 + \Pbr{\frac{1}{\A},\Re\Gr^R}\right)
C\left(X^\mu - \frac{1}{A}\frac{\partial \Re\Gr^R}{\partial
p_{\mu}}, p^\mu+\frac{1}{A}\frac{\partial \Re\Gr^R}{\partial
X_{\mu}}\right).
\end{eqnarray}
%
 Eq. (\ref{new-kin-eq}) is the key equation of
our work.  The collision term $C^{\rm NL}$ is expressed here in
terms of shifted variables. 
 Note that the specific 8-phase-space memory in $C^{\rm
NL}$ should not be confused with  the memory effects resulting
from internal structure of self-energy diagrams with more than two
vertices. In order to distinguish between these two effects we
call the non-locality in Eq. (\ref{new-kin-eq}), as delays
(positive or negative)
in 8-phase-space rather than the memory.

To verify that the NL kinetic equation  Eq. (\ref{new-kin-eq}) is
equivalent to the KB one,
 including
second-order gradient terms, let us expand the collision term
$C^{\rm NL}$ in small quantities
%
\begin{eqnarray}
\label{delX}
\delta X^\mu = (\delta \tau,\delta {\bf x})=
\frac{1}{A}\frac{\partial \Re\Gr^R}{\partial p_{\mu}}, \quad
\label{delta-p}
 \delta p^\mu = \frac{1}{A}\frac{\partial
\Re\Gr^R}{\partial X_{\mu}}.
\end{eqnarray}
%
Then
%
\begin{eqnarray}
\label{Cnl-exp}
C^{\rm NL} &=&
\left(1 + \Pbr{\frac{1}{\A},\Re\Gr^R}\right)
\left(C +\frac{1}{\A}\Pbr{C,\Re\Gr^R}\right) + O(\partial_X^3)
\cr
&=&
\vphantom{\frac{1}{\A}}
C +
\Pbr{C/\A,\Re\Gr^R} + O(\partial_X^3).
\end{eqnarray}
%
Here we have taken into account that $C\propto O(\partial_X)$. The
resulting Poisson bracket $\Pbr{C/\A,\Re\Gr^R}$ combines with the
Poisson bracket on the l.h.s., cf. Eq. (\ref{BM-subst}), resulting
in $\Pbr{\Ldt,\Re\Gr^R}$, as it stands in the KB kinetic equation
(\ref{keqk1}). Thus, indeed, Eq. (\ref{new-kin-eq}) coincides with
the KB equation (\ref{keqk1}) within the approximation 
including second-order terms in space-time gradients. The
roughening of this approximation by neglecting the second-order terms 
(i.e., if one puts $C^{\rm NL}=C$), returns
us to the BM equation.

The consistency condition related to Eq. (\ref{new-kin-eq}) can be
rewritten as
\begin{equation}
\label{invarJk-new} e \mbox{Tr} \int \dpi{p} \left\{
-\partial_\mu \left[  \frac{\partial
\Re\Se^R}{\partial p^{\mu}}F
+
\Re\Gr^R \frac{\partial (\Gm {F}/{\A})}{\partial p_{\mu}} \right]
+ C^{\rm NL} +O(\partial_X^3)\right\} = 0 ,
\end{equation}
cf. Eq. (\ref{invarJk}). 
Thus,  the precision of
conservations of the Noether baryon number and the
energy--momentum in the NL equation is one order in  space-time
gradients higher than that in the BM equation.  Indeed, the effective
BM current differs from the Noether one in terms of the first-order
gradients  whereas the effective current that follows from the
NL kinetic equation differs 
only in
second-order gradients. The latter takes place because the NL
equation is equivalent to the KB equation including
$O(\partial_X^2)$ terms. Let us recollect that Noether
currents are exactly conserved for the KB choice \cite{KIV01}.

Within the range of formal applicability, the NL 
equation is equivalent to the KB and BM forms
discussed above. However,
in practical calculations one frequently needs to  apply the
kinetic approach beyond the scope of its applicability. For
instance, such a problem arises when one  describes the initial
stage of heavy ion collisions. We do not know how good the quantum
kinetic equations are beyond the region of their formal
applicability, but we certainly wish to respect conservation 
laws. From this point of view the KB form is certainly preferable. As
the conservations are exact, we can still use the gradient
approximation, relying on a minor role of this rather short
initial stage of heavy ion collisions in the total evolution of a
system. Exact conservation laws allow us to keep control of
numerical codes. However, as we have mentioned, 
the efficient test-particle method is
not applicable for solution of the KB equation. 
At the same time, for the BM
kinetic equation the test-particle method is already available
\cite{Cass99,Leupold}, for the price that it deals with the 
approximately conserved Noether current.

When the collision term $C$ is large, the neglect of the $C/A$ term
in the r.h.s. of the relation (\ref{BM-subst}) is a bad
approximation. Please, recollect that this approximation is in the
basis of the BM approximation. Therefore, the nonlocal form of the
kinetic equation which we present here is certainly preferable.
Conservations of the Noether current and energy-momentum are more
accurate here than in the BM form, since they are closer to exact
conservations of the KB form.\footnote{Deriving
(\ref{new-kin-eq}), we only added higher (than second order) space-time
derivatives to
   equation in KB choice (\ref{keqk1}).}
At the same time the test-particle method is applicable to the
NL equation. 
In the NL case the set of equations for evolution of
parameters of the test particles between collisions is the same as
in the BM case, see Eqs. (\ref{x-dot})--(\ref{e-dot}).  The only
difference with the BM case is that collisions of test particles
occur with certain time (and space) delay (or advance) as compared
with the instant of their closest approach to each other. Because
of this circumstance the effective BM current
(\ref{BM-eff-currentk}) is no longer locally conserved.
Indeed, integrating $C^{\rm NL}$ over 4-momentum and taking into
account relation (\ref{Cnl-exp}), we arrive at
%
\begin{eqnarray}
\label{C-loc.cons}
\mbox{Tr} \int \dpi{p}C^{\rm NL} &=&
\mbox{Tr} \int \dpi{p} \Pbr{C/\A,\Re\Gr^R} + O(\partial_X^3)
\cr
&=&
\partial_\mu \left(-\mbox{Tr} \int \dpi{p}
\frac{C}{\A} \frac{\partial\Re\Gr^R}{\partial p_{\mu}}\right)
+ O(\partial_X^3).
\end{eqnarray}
%
Precisely this contribution makes the KB effective current
(\ref{KB-eff-currentk}) from the corresponding BM current
(\ref{BM-eff-currentk}). This nonconservation does not prevent us
from application of the test-particle method. Since collisions are
nonlocal in the time--space, we need only global conservation of the
BM effective current, which in fact takes place
%
\begin{eqnarray}
\label{C-tot.cons}
\mbox{Tr} \int \di^4 X \ \dpi{p}C^{\rm NL} &=&
\int \di^4 X \partial_\mu \left(-\mbox{Tr} \int \dpi{p}
\frac{C}{\A} \frac{\partial\Re\Gr^R}{\partial p_{\mu}}\right)
+ O(\partial_X^2) = O(\partial_X^2).
\end{eqnarray}
%
Now this conservation is approximate, $O(\partial_X^2)$ terms are
dropped. Nevertheless, this precision is better than that
resulting from the formal accuracy of the kinetic
equations, where already  $O(\partial_X)$ terms are dropped in the
approximately conserving quantities.

Actually we could  modify the NL equation in such a way that it exactly
fulfills  global  conservation  of the effective
charge, which may be important for the test-particle method. 
Indeed, we recognize that to the first-order gradients 
the pre-factor
$J=\left(1 + \Pbr{\frac{1}{\A},\Re\Gr^R} \right)$ in  Eq.
(\ref{new-kin-eq}) is the Jacobian $J$ of the transformation from
$\di^4 p \di^4X$ to the
 $\di^4 \widetilde{p} \di\widetilde{X}$ integration 
expressed in terms of shifted variables
$\widetilde{p}^{\mu} =p^{\mu}+\delta p^{\mu}$ and 
$\widetilde{X}^{\mu} =X^{\mu}-\delta X^{\mu}$. 
By adding terms $\delta J$ of higher (than $O(\partial_X)$) orders
to $J$, we 
do not violate the nature of the NL approximation. Then  
$\delta J$ can be tuned in such a way that $J+\delta J$ becomes the
exact Jacobian of the transformation and hence the BM effective
current turns out to be exactly globally conserved:  
\begin{eqnarray}
\label{C-tot.cons.exact} 
\mbox{Tr} \int \di^4 X \ \dpi{p}
\frac{J+\delta J}{J}
C^{\rm NL}=
\mbox{Tr} \int \di^4 \widetilde{X} \ \dpi{\widetilde{p}} C
(\widetilde{X}, \widetilde{p}) =0.
\end{eqnarray}
However the explicit expression for $\delta J \propto O(\partial_X^2)$ is
very cumbersome. In order to keep the formalism transparent, we avoid
this extra complication.

\subsection{Time delays and causality}\label{phys}

The NL kinetic equation (\ref{new-kin-eq}) contains the delayed
collision term and the BM particle drift term including the drag
and the back flow contributions.
The extra Poisson-bracket term
$\Pbr{C/\A,\Re\Gr^R}$
 which in the KB equation had a poorly defined physical
meaning is now hidden in the 8-phase-space delays
of the collision term thus acquiring a new physical
interpretation. The space-time 
part of the retardation/advance (\ref{delX}) is expressed as
%
\begin{eqnarray}\label{st-ret}
\delta X^{\mu} =
\frac{1}{2}B^{\mu}-\frac{1}{\Gamma}
\left(v^{\mu}  - \frac{\partial
\Re\Se^R}{\partial p_{\mu}}\right),
\end{eqnarray}
%
where we used Eq. (\ref{Asol}) and the definition of the flow
spectral
function (\ref{B-j}). There are two
contributions to the $\delta X^{\mu}$. The first term
is proportional to the same $B^{\mu}$
function that determines the
trajectories of the test particles between collisions,
see Eqs. (\ref{x-dot})--(\ref{e-dot}). The second
term $\left(v^{\mu}  - \partial \Re\Se^R/\partial
p_{\mu}\right)/{\Gamma}$ relates to collisions: $1/\Gamma \sim
1/(u\sigma \rho)$ in the gas approximation, where $u$ is the
particle velocity,  $\rho$ is the density of the medium and
$\sigma$ is the cross-section.

Note that the time delay in the collision term has no definite
sign (see discussion in the next subsection). Therefore, one could
naively suppose that NL kinetic equation violates causality. Here
the following comment is in order.  The initial Dyson equations
for non-equilibrium Green functions are causal. The apparent
problem appears already at the moment when one chooses $t=(t_1
+t_2)/2$ and $\Delta t =t_2 -t_1 $ as independent variables (where
$t_2
>t_1$ due to causality) to perform then the Wigner
transformation. Although approximations have not yet been done, in
new variables the Green function $G^R (t,\Delta t)$ contains an
information from the future, see Fig. 19 of Ref. \cite{Greiner}.
The problem would completely disappear, if one had used $t_1$ and
$\Delta t =t_2 -t_1 $ as independent variables. Thus, working in
$t=(t_1 +t_2)/2$ and $\Delta t =t_2 -t_1 $ variables we 
incorporate some information from the future. Otherwise our
results would be incompatible with those obtained from the
original
 Dyson equations.
However, this fact does not contradict causality, since the Dyson
equations are causal and approximations have not yet been done.
The necessity to include the information from the future is
clearly seen, e.g., from analysis of the $\Phi$-diagrams
containing more than two
 vertices. The  vertices depend on different time variables.
 These dependencies reveal themselves in the  above mentioned memory
 effects  in the gradient expanded kinetic equations.

The apparent acausality  disappears after  the first-order
gradient expansion, and it does not manifest in the KB and BM
forms of the kinetic equations since all variables there depend
locally on the single time $t$. However via corresponding Poisson
bracketed terms  these equations contain an information on the
mentioned non-local effects, which were  originally present.
Constructing (\ref{new-kin-eq}) we re-grouped some of these terms
hiding them in the collision term at the price that the apparent
problem again appeared. 
However, the NL kinetic equation can be read from the right to the left in
such a way that the collision term in future is determined by the
drift term at present. This way the NL equation is again completely
causal. 

As we will see in the next section  the
new NL form of the kinetic equation is  convenient to discuss
the so called negative time delays which have been experimentally
observed in semiconductors, see \cite{Axt} and refs. therein. The
quantum kinetic delay effects reflect fundamental physical
limitations which follow from basic uncertainty principles for the
time--energy and the coordinate--momentum. The finite duration
(positive or negative) implies that the final state of the system
is not entirely determined until the process is completed. The
wave-like nature of excitations  indicates that the interactions
between waves are interference phenomena which have finite
duration on a scale of at least one period of the interacting
waves. Collisions cannot be considered to be completed at times
shorter than these periods. This may result in that a collision in
progress can be reversed at some conditions. Quantum nature of
collisions is reflected in the formalism of quantum kinetics
contrary to the Boltzmann kinetics dealing with free particle
propagation between instant collisions.

\subsection{Physical meaning of time delays and advances}

In order to consider the variable shifts in the collision term in more
detail, let us present the time delay as a sum of two quantities
\begin{eqnarray}\label{tdel}
\delta \tau =\delta \tau_{\rm drift}+\delta \tau_{\rm col}\equiv
\frac{B_0}{2}- \frac{Z^{-1}}{\Gamma},
\end{eqnarray}
where
$Z=\left(v_{0}  - \partial \Re\Se^R/\partial p^{0}\right)^{-1}$
is the normalization factor which usually appears in
description of quasiparticle effects. In contrast to the
quasiparticle case, here $p_0$ and $\vec{p}$ are independent
variables (not connected by any dispersion relation).
The $B_0$ function can be formally expressed as
\begin{eqnarray}\label{ph-sh}
B_0 =2\frac{\partial \delta}{\partial p_0},\,\,\,\,
\mbox{tan}\delta =-\frac{\Gamma}{2M}.
\end{eqnarray}
 The latter ratio is a measure of proximity of a virtual particle with
 given $p_0$ and $\vec{p}$  to the mass-shell, see Eq. (\ref{off-shellness}) above.

The value $\delta \tau_{\rm drift}=B_0/2$ has the meaning of the
drift delay time, $\delta \tau_{\rm col}=-Z^{-1}/\Gamma$, of the
collision delay time, and $\delta \tau$, of the effective total
delay time. In the virial limit these time delays coincide with
those studied in \cite{Bosanas,DP}.
In particular, for quasiparticles $Z>0$ and $M\rightarrow 0$,
and therefore  we obtain
 $B_0 =Z A =2\pi Z \delta (M)>0$ and hence
 $\delta \tau_{\rm drift}$ is positive delay,
and $\delta \tau_{\rm col}$ is
negative delay at arbitrary densities.
To further clarify the meaning of these delays let us consider several
examples.

\begin{itemize}

\item
{\em $\pi N\Delta$ system}

If we consider propagation of a resonance, e.g., $\Delta$ isobar,
in the virial limit of very small densities, $\delta$ has the
meaning of the $\pi N$ phase shift. Thus in the virial limit,
$B_0$
 of the resonance relates to the energy variation
of the  scattering phase shift ($\delta_{33}^{\pi N}$) of the
channel coupled to the resonance, cf. \cite{Weinhold}. Due to the
Galilean invariance the vacuum Green function and self-energy
depend only on the energy in the $\pi N$ center-of-mass frame and
$B_0 $ is recovered from the known dependence $\delta_{33}^{\pi N}
(E_{\rm c.m.})$. As it follows from this dependence, $B_0 >0$ and
hence $\delta\tau_{\rm drift} >0$ (delay). There also appears  a
negative delay $\delta\tau_{\rm col}<0$ (for $Z>0$), since namely
during the $Z_{\Delta}^{-1}/\Gamma_{\Delta}$ time  the $\Delta$
resonance state forms  the intermediate state of the $\pi N$ pair
(causing an advance), compare with Ref. \cite{DP}. The $\delta
\tau$ value in Eq. (\ref{tdel}) has no definite sign. Its sign
depends on which time is dominant, $\delta\tau_{\rm drift}$ or
$\delta\tau_{\rm col}$. If we consider the pion propagation, $B_0
=B_0^{\pi}$ and in the virial limit $\delta$ has the meaning of
the in-medium $\Delta$ -- nucleon hole phase shift. For the
nucleon, $B_0^{N}$ is related to $\delta_{\pi \Delta}$.

\item
{\em Wigner resonances}

For a Wigner nonrelativistic resonance, when
$\Re\Sigma^R$ and $\Gamma$ are independent of $p$, we
have $Z=1$. Then the value $\delta \tau$ in Eq. (\ref{tdel}) becomes
\begin{eqnarray}\label{res-vic}
\delta \tau = -\frac{M^2 -\Gamma^2/4}{\Gamma (M^2 +\Gamma^2/4 )},
\end{eqnarray}
which precisely coincides with the result Ref. \cite{DP}.
As it follows from (\ref{res-vic}), in the vicinity of the
resonance ($M\rightarrow 0$) $\delta \tau$ is positive (delay) and
outside the resonance region ($M^2\gg \Gamma^2$) $\delta \tau$ is
negative (advance).
 If the
interaction is attractive, the particle spends  a shorter time in the
interaction zone than that would be, if it moved with unchanged
asymptotic velocity. This case corresponds to a delay. If the
interaction is repulsive, the particle vise versa spends a longer
time in the interaction zone, that corresponds to a negative
delay. Thus,
close to the Wigner resonance we may speak about an effective
attraction and outside the resonance region, about   an effective
repulsion.

\item
{\em Soft photon radiation}

Let us consider another example
 \cite{KV96} of the photon radiation in
the propagation of a source charge (say a proton) in neutral
matter (e.g. the neutron one).
 The source particle looses the
memory about collisions which have happened long ago (after a time
$\tau \gg |\delta\tau_{\rm col}|$ has past). The advance in the
collision time can be understood in terms of multiple scatterings.
The latter act coherently for $\tau \lsim |\delta\tau_{\rm
col}|\sim 1/\Gamma$. Indeed, for energies of  the radiating
quantum corresponding to the wavelength $\lambda_{\gamma}
=2\pi/\omega_{\gamma} \gsim |\delta\tau_{\rm col}|$ the source
particle simultaneously feels all rescatterings which it had and
also will have on the time scale $|\delta\tau_{\rm col}|$. This
results in the well known Landau-Pomeranchuk-Migdal effect
\cite{LPM} experimentally proven at the Stanford Linear
Accelerator Center \cite{Antony}. For large photon energies,
$\lambda_{\gamma}$ is small and radiation occurs on a single
center (neutron), whereas soft radiation occurs coherently on many
centers. Thus appearance of the time advances $\delta\tau_{\rm
col}$ in this example is associated with the
 multiple rescattering effects.

\end{itemize}


In the above examples   $Z$ factors were assumed to be positive.
However, $Z$ can be negative in some energy-momentum regions far
from the mass-shell, cf. \cite{MSTV90}.  Change of the sign of $Z$
is reflected in the respective sign change in $\delta \tau_{\rm
col}$.

Moreover, interactions lead to the change in the generalized group
velocity 
\begin{eqnarray}
\label{Groupv}
\vec{u}_{\rm group}=Z\left(\vec{v}+\frac{\partial
\Re\Se^R}{\partial \vec{p}}
 \right)
\end{eqnarray}
that reflects in the time-space delays. To illustrate it let us
consider an example of the pion propagation in the
isospin-symmetric nuclear matter. For the sake of simplicity, we
consider the  pion propagation in a region close to the
$\Delta$-resonance and assume that the pion interacts with the
nucleons only through the $\pi N\Delta$-coupling $f_{\pi
N\Delta}$. Then the real part of the pion self-energy becomes
\cite{MSTV90}
\begin{eqnarray}
\Re\Sigma_{\pi}\simeq \frac{\alpha_0 f^2_{\pi N\Delta}\vec{k}^2
\rho_N \omega_{\Delta}[\omega^2 -\omega_{\Delta}^2]}{(\omega^2
-\omega_{\Delta}^2)^2 +\Gamma^2 \omega_{\Delta}^2}.
\end{eqnarray}
Here $\omega_{\Delta}=m_{\Delta}-m_N$, $\rho_N$ is the nucleon
matter density, $\alpha_0 =const \sim 1$. From here we can find
$Z_{\pi}$-factor. For $\omega\rightarrow \omega_{\Delta}$ we
obtain
\begin{eqnarray}
Z_{\pi}\rightarrow  \left[2\omega_{\Delta}  - \frac{2\alpha_0
f^2_{\pi N\Delta}\vec{k}^2 \rho_N
}{\Gamma_{\Delta}^2}\right]^{-1},\quad \vec{u}_{\rm
group}\rightarrow 2\vec{k} Z
\end{eqnarray}
resulting in an increase of the group velocity in the resonance
region. Here it is worthwhile to note that for light in a
dielectric medium there appears increase in the group velocity for
packets with resonant frequencies \cite{Jackson}.

\subsection{Space and energy-momentum shifts}

As follows from Eq. (\ref{delX})  the shift of the space variables
is
\begin{eqnarray}\label{delX1}
\delta \vec{x} =\frac{1}{A}\frac{\partial \Re\Gr^R } {\partial
\vec{p}} =\delta \vec{x}_{\rm drift}+\delta \vec{x}_{\rm col
}=\frac{1}{2}\vec{B}+\vec{u}_{\rm group}\delta\tau_{\rm col}.
\end{eqnarray}
%
%
We can further express the spatial shift (\ref{delX1}) in terms of
time delays (\ref{tdel}) and velocity ${\bf {\dot x}}$  of a test
particle on its trajectory,\footnote{Here for briefness  we omit
the unnecessary subscript $i$ of the test particle.} cf. Eq.
  (\ref{x-dot}),
\begin{eqnarray}
\label{delX2}
\delta \vec{x} ={\bf {\dot x}} \delta \tau_{\rm drift}
+\vec{u}_{\rm group}\delta\tau_{\rm col}.
\end{eqnarray}
Eq. (\ref{delX2}) demonstrates that to a delayed collision the
test particle moves along its trajectory. Therefore, the drift
time delay $\delta \tau_{\rm drift}$ unambiguously results in a
definite space shift  ${\bf {\dot x}} \delta \tau_{\rm drift}$.
The collision itself is associated with an additional time delay
$\delta\tau_{\rm col}$, which implies that the collision is not
instant, as it is treated in the kinetic equation, but requires
certain time for complete decoupling from intermediate states
(e.g., the pion spends some time in the intermediate
$\Delta$--nucleon-hole state, a soft photon requires certain time
to be formed in multiple collisions of the proton with neutrons).
Therefore, this additional delay gives rise to an additional shift
$\vec{u}_{\rm group}\delta\tau_{\rm col}$ of the particle with
respect to its ``collisionless'' trajectory (\ref{x-dot}).

Since during time-delay interval $\delta \tau$ a
particle propagates in the mean field $\Re\Se^R$, it
also changes its energy and momentum.
The energy--momentum shift (\ref{delta-p}) can be represented in a
similar form
\begin{eqnarray}
\label{delta-p2} \delta p^\mu={\dot p}^\mu \delta \tau_{\rm drift}
-Z\frac{\partial\Re\Se^R}{\partial X_{\mu}} \delta\tau_{\rm col},
\end{eqnarray}
where ${\dot p}^\mu=({\dot E},{\bf {\dot p}})$ is the rate of change
of the 4-momentum of a test particle on its trajectory, cf. Eqs.
(\ref{p-dot}) and (\ref{e-dot}). We again distinctly see two
contributions to $\delta p^\mu$. The first one is associated with the
energy--momentum change during the path of the particle to
a delayed collision. The second one is related to an intrinsic
nonlocal character of the collision itself.
Thus, we see that all shifts in space and energy--momentum variables
are consistent with time delays discussed above.

The energy shift can be expressed in the alternative form
\begin{eqnarray}
\delta p_0
=
\frac{M^2 -\Gamma^2/4}{M^2 +\Gamma^2/4}\left(\frac{\partial_t
\Re\Se^R} {\Gamma}\right)-\Re\Gr^R\partial_t \Gamma,
\end{eqnarray}
where we used Eq. (\ref{Asol}). In the case of the Wigner
resonance, the first term on the r.h.s. of this relation is
$-\delta \tau\cdot \partial_\tau \Re\Se^R $. Thus, the
energy-shift scale is $\sim |\Re\Se^R \cdot \delta \tau
|/\tau_{\rm ch}$, where $\tau_{\rm ch}$ is the characteristic time
of the self-energy variation. In the virial limit $\Re\Se^R =4\pi
\rho \Re F(0)$, where $F(0)$ is the forward scattering amplitude
in the energetic normalization, $\rho$ is the density of the
medium. Then we obtain
\begin{eqnarray}
\delta p_0 \sim -4\pi \Re F(0) \delta \tau \cdot \partial_t \rho,
\end{eqnarray}
thus relating the value $\tau_{\rm ch}$ to the variation of the
density. In the kinetic region the $\tau_{\rm ch}$ scale is the
collisional time, $\tau_{\rm ch}\sim |\delta\tau_{\rm col}|$, and
hence the correction $\delta p_0$ can be not small. In the
region of the validity of the hydrodynamics the density varies on
a much longer scale than the collisional one, and hence
the correction $\delta p_0$ is small but still does not vanish.
When the global equilibrium is reached this correction disappears.
\\[5mm]

Since the time delay occurs in all the quantities
entering the collision term, and the value of the
$\delta \tau$ delay is finite even in the case of global
equilibrium (e.g., see Eq. (\ref{res-vic})), the thermodynamic
quantities, such  as the pressure, energy, etc., also acquire
corrections proportional to this time delay.

 At the general level, the
relation of thermodynamic properties to the time delays can be
understood in terms of ergodicity. The system spends certain time in a
certain phase-space region, which is proportional to the density of states
in that region. Change in this time corresponds to the change in
the density of states and thereby  to the  change in thermodynamic
properties, cf. \cite{DP}. We have generalized this statement to
the unified 8-phase-space.

Another important feature of the kinetic description is
approaching to thermal equilibrium during the evolution of a
closed system. In terms of transport theory it means that the
H-theorem takes place for transport equations, which implies
construction of the entropy related to the  kinetic equation.
These problems have been considered both for the BM equation and
for the KB equation \cite{IKV99,IKHV00,KIV01,IKV03}. Since the NL
form of the quantum kinetic equation (\ref{new-kin-eq}) is
equivalent to the KB equation including second-order terms in
space--time gradients, the NL entropy differs from 
the KB entropy only in the second-order gradient terms.
This reduces the problem of the H-theorem for the new NL kinetic
equation to the already solved problem for the KB equation
\cite{IKV03}.

\section{Summary}\label{Summary}

We have considered properties (advantages and disadvantages) of
available formulations of the off-shell kinetic equation, i.e.
those in the Kadanoff--Baym and Botermans--Malfliet forms, as well
as a new nonlocal form which we derived in this paper. 
Within the
range of formal applicability, all these forms are equivalent to
each other. Under this applicability range we mean states of a
system which are close to equilibrium, where the space--time
gradient expansion is strictly applicable. 
In particular, it
implies that conservation laws of Noether currents and
energy-momentum are satisfied at least to the extent of validity
of the dynamics, i.e. up to zero-order gradients. Note that
these conservation laws are even exactly fulfilled for the
Kadanoff--Baym form \cite{KIV01}.
However, these
equations are frequently applied beyond the scope of their
applicability, e.g., when the collision term is large. Such a
situation takes place, e.g., in description of the initial stage
of heavy-ion collisions. 
In such situations we certainly wish to respect conservation laws 
which allow us to keep control of numerical codes. From this point of
view the KB form is certainly preferable. As 
the conservations are exact, we can still use the gradient
approximation, relying on a minor role of this rather short
initial stage of heavy ion collisions in the total evolution of a
system. However, the efficient test-particle method is
not applicable for solution of the Kadanoff--Baym equation.

The nonlocal kinetic equation differs from the Kadanoff--Baym one
only in third-order gradient terms, while the
Botermans--Malfliet form, in the second order.
Due to this the nonlocal form of the kinetic equation
is certainly preferable as compared with the Botermans--Malfliet one,
since it more accurately conserves the Noether currents and the
energy-momentum. Deviations from the Noether quantities appear
only in the second gradient order, as compared with first order for
the Botermans--Malfliet equation. 
In addition, the nonlocal form allows
application of the test-particle method very similarly to that has
been done for the Botermans--Malfliet form \cite{Cass99,Leupold}.
Presently, there are numeric schemes dealing with delayed
(advanced) collisions of the test particles \cite{DP,Spicka98}.
Thus, the nonlocal form of the quantum kinetic equation is a
reasonable compromise between the Kadanoff--Baym and
Botermans--Malfliet equations, which is practical like the
Botermans--Malfliet form and at the same time reasonably preserves
conservations of Noether quantities.

Physical meaning of delays and advances in 8-dimensional phase space
appearing in the nonlocal form of the kinetic equation is clarified. 
It is demonstrated that in addition to the practical utility the
nonlocal form provides us with clear physical interpretation of
various terms in the Kadanoff--Baym equation and also 
allows us to consider delicate physical phenomena, which
are beyond the scope of the Botermans--Malfliet approximation.

\section*{Acknowledgments}

We are grateful to  J. Knoll for valuable remarks and constructive
criticism. Also we acknowledge   E.E. Kolomeitsev for fruitful
discussions. This work was supported by the Deutsche
Forschungsgemeinschaft (DFG project 436 RUS 113/558/0-3), the
Russian Foundation for Basic Research (RFBR grant 06-02-04001
NNIO\_a), and Russian Federal Agency for Science and Innovations
(grant NSh-3004.2008.2).

\appendix

\section{Matrix Notation} \label{Contour}

In calculations that apply the Wigner transformations, it is necessary to
decompose the full contour into its two branches---the {\em time-ordered} and
{\em anti-time-ordered} branches. One then has to distinguish between the
physical space-time coordinates $x,\dots$ and the corresponding contour
coordinates $x^{\cal C}$ which for a given $x$ take two values
$x^-=(x^-_{\mu})$ and $x^+=(x^+_{\mu})$ ($\mu\in\{0,1,2,3\}$) on the two
branches of the contour (see figure 1). \\
\begin{center}
\vspace*{0.1cm}
\contourxy\\[0.5cm]
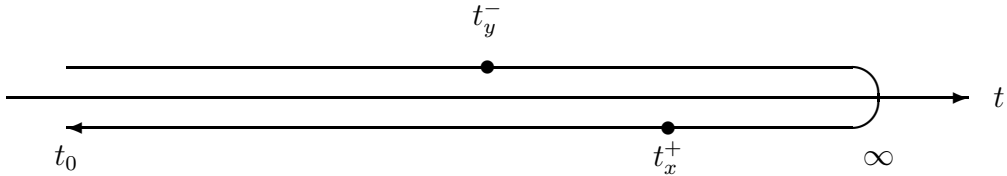
{\small Figure 1: Closed real-time contour with two external
  points $x,y$ on the contour.}
\end{center}

Closed real-time contour integrations
can then be decomposed as
%
\begin{eqnarray}
\label{C-int}
\oint\di x \dots =\int_{t_0}^{\infty}\di x\dots
+\int^{t_0}_{\infty}\di x\dots
=\int_{t_0}^{\infty}\di x\dots -\int_{t_0}^{\infty}\di x\dots,
\end{eqnarray}
%
where only the time limits are explicitly given.  The extra minus sign of the
anti-time-ordered branch can conveniently be formulated by a $\{-+\}$
``metric'' with the metric tensor in $\{-+\}$ indices
%
\begin{eqnarray}
\label{sig}
\left(\sigma^{ij}\right)&=&
\left(\sigma_{ij}\vphantom{\sigma^{ij}}\right)=
{\footnotesize\Big(\begin{array}{cc}1&0\\[-3mm]
0& -1\end{array}\Big)}
\end{eqnarray}
%
which provides a proper matrix algebra for multi-point functions on the
contour with ``co''- and ``contra''-contour values.  Thus, for any two-point
function $F$, the contour values are defined as
%
\begin{eqnarray}\label{Fij}
F^{ij}(x,y)&:=&F(x^i,y^j), \quad i,j\in\{{\scriptstyle -\,,\, +}\},\quad\mbox{with}\cr
F_i^{~j}(x,y)&:=&\sigma_{ik}F^{kj}(x,y),\quad
F^i_{~j}(x,y):=F^{ik}(x,y)\sigma_{kj}\cr
F_{ij}(x,y)&:=&\sigma_{ik}\sigma_{jl}F^{kl}(x,y),
\quad\sigma_i^k=\delta_{ik}
\end{eqnarray}
%
on the different branches of the contour. Here summation over repeated indices
is implied. Then contour folding of contour two-point functions, e.g., in Dyson
equations, simply becomes
%
\begin{eqnarray}\label{H=FG}
H(x^i,y^k)&=&H^{ik}(x,y)= \left[ F\otimes G \right]^{ik}
\cr
&\equiv& \oint\di z F(x^i,z)G(z,y^k)
=\int\di z F^i_{~j}(x,z)G^{jk}(z,y)
\end{eqnarray}
%
in the matrix notation.

Due to the change of operator ordering, genuine multi-point functions are, in
general, discontinuous, when ever two contour coordinates become identical. In
particular, two-point functions like $\ii F(x,y)=\left<\Tc {\widehat
    A(x)}\medhat{B}(y)\right>$ become\footnote{Frequently used
  alternative notation is $F^{<}=F^{-+}$ and $F^{>}=F^{+-}$.}
%
\begin{eqnarray}\label{Fxy}
\hspace*{-0.5cm}\ii F(x,y) &=&
\left(\begin{array}{ccc}
\ii F^{--}(x,y)&&\ii F^{-+}(x,y)\\[3mm]
\ii F^{+-}(x,y)&&\ii F^{++}(x,y)
\end{array}\right)=
\left(\begin{array}{ccc}
\left<{\cal T}\medhat{A}(x)\medhat{B}(y)\right>&\hspace*{5mm}&
\mp \left<\medhat{B}(y)\medhat{A}(x)\right>\\[5mm]
\left<\medhat{A}(x)\medhat{B}(y)\right>
&&\left<{\cal T}^{-1}\medhat{A}(x)\medhat{B}(y)\right>
\end{array}\right),
\end{eqnarray}
%
where ${\cal T}$ and ${\cal T}^{-1}$ are the usual time and anti-time ordering
operators.  Since there are altogether only two possible orderings of the two
operators, in fact given by the Wightman functions $F^{-+}$ and $F^{+-}$,
which are both continuous, not all four components of $F$ are independent. Eq.
(\ref{Fxy}) implies the following relations between nonequilibrium and usual
retarded and advanced functions
%
\begin{eqnarray}\label{Fretarded}
F^R(x,y)&=&F^{--}(x,y)-F^{-+}(x,y)=F^{+-}(x,y)-F^{++}(x,y)\nonumber\\
&:=&\Theta(x_0-y_0)\left(F^{+-}(x,y)-F^{-+}(x,y)\right),\nonumber\\
F^A (x,y)&=&F^{--}(x,y)-F^{+-}(x,y)=F^{-+}(x,y)-F^{++}(x,y)\nonumber\\
&:=&-\Theta(y_0-x_0)\left(F^{+-}(x,y)-F^{-+}(x,y)\right),
\end{eqnarray}
%
where $\Theta(x_0-y_0)$ is the step function of the time difference.  The
rules for the co-contour functions $F_{--}$ etc. follow from Eq. (\ref{Fij}).

For slightly inhomogeneous and slowly evolving systems, the degrees of freedom
can be subdivided into rapid and slow ones. Any kinetic approximation is
essentially based on this assumption.  Then for any two-point function
$F^{ij}(x,y)$, one separates the variable \mbox{$\xi =(t_1-t_2,
  \vec{r_1}-\vec{r_2})$}, which relates to rapid and short-ranged microscopic
processes, and the variable $X= \frac{1}{2}(t_1+t_2,\vec{r_1}+\vec{r_2})$,
which refers to slow and long-ranged collective motions. The Wigner
transformation, i.e.  the Fourier transformation in four-space difference
$\xi=x-y$ to four-momentum $p$
%
\begin{equation}
\label{W-transf}
F^{ij}(X;p)=\int \di \xi e^{\ii p\xi}
F^{ij}\left(X+\xi/2,X-\xi/2\right),
\quad\quad\quad i,j\in\{-+\}
\end{equation}
%
where $X=(t,{\bf x})$,
leads to the corresponding Wigner densities in four-phase-space.
The gradient expansion converts the Wigner transformation of any convolution of
two-point functions into a product of the corresponding Wigner functions
plus higher order gradient terms
%
\begin{eqnarray}\label{CK}
\int \di \xi e^{\ii p \xi}
\left(
\int \di z f(x,z) \varphi(z,y)
\right) &=&
\left(
\exp\left[\frac{\ii\hbar}{2}\left(\partial_p \partial_{X'}-
\partial_X \partial_{p'}\right)\right] f(X, p) \varphi (X',p')
\right)_{p'=p,X'=X}\\ 
\label{g-rule2}
&\simeq&
f(X,p) \varphi(X,p) +
\frac{\ii\hbar}{2}
\Pbr{f(X,p), \varphi(X,p)},
\end{eqnarray}
%
where the first order terms are given by Poisson brackets
%
\begin{equation}
\label{[]}
\Pbr{f(X,p) , \varphi(X,p)} =
\frac{\partial f}{\partial p^{\mu}}
\frac{\partial \varphi}{\partial X_{\mu}}
-
\frac{\partial f}{\partial X^{\mu}}
\frac{\partial \varphi}{\partial p_{\mu}}
\end{equation}
%
here in covariant notation.  We would like to stress that the
smallness of the $\hbar\partial_X \cdot\partial_p$ comes solely
from the smallness of space--time gradients $\partial_X$, while
momentum derivatives $\partial_p$ are not assumed to be small.
This point is sometimes incorrectly treated in the literature.

\end{fmffile}

\end{document}